# Large Nonlinear Kerr Angle in non-Centrosymmetric Fe/AlGaAs (001) Heterostructure

Haibin Zhao[1], Diyar Talbayev[1], Gunter Luepke[1], Aubrey Hanbicki[2], Connie Li[2], and Berend Jonker[2]

[1]The College of William and Mary, Williamsburg, VA, 23187
[2]Naval Research Laboratory, Washington, DC, 20375

Abstract

A large nonlinear magneto-optical effect is observed in a non-centrosymmetric Fe/AlGaAs (001) heterostructure. This effect is a direct consequence of interference between second-harmonic optical waves of magnetic and crystallographic origin, generated at ferromagnetic Fe interface and bulk AlGaAs, respectively. The longitudinal nonlinear Kerr rotation is measured to be 1.6º along the [1-10] hard axis, about two orders of magnitude stronger than the linear equivalent. The rotational second-harmonic signal shows large magnetic contrast along all the in-plane directions, demonstrating a high sensitivity to the magnetization of an anisotropic interface in the longitudinal geometry.

## Introduction

The nonlinear magneto-optical Kerr effect is sensitive to surface and interface magnetization in centrosymmetric thin films [1,2] and multilayers [3] due to the fact that the crystallographic bulk contribution to second harmonic generation (SHG) is forbidden in centrosymmetric materials in electric dipole approximation. Moreover, the magnetization does not break the inversion symmetry in the bulk but can lower the surface and interface symmetry, leading to magnetization-induced SHG (MSHG) only from the latter [4]. In this case, magnetization-induced components in MSHG are comparable in magnitude with components of non-magnetic origin, which leads to high magnetic asymmetry and large nonlinear magneto-optical Kerr angle [5]. This intrinsic sensitivity, however, has so far only been shown for centrosymmetric structures. In this paper, we will demonstrate a large nonlinear magneto-optical effect from buried Fe interface in a noncentrosymmetric Fe/AlGaAs hybrid structure. The nonlinear magneto-optical Kerr angle is measured to be 1.6º, which is about two orders of magnitude stronger than the linear equivalent.

## Theory

For a particular magneto-optical configuration, the optical second-harmonic polarization **P** (2ω) of the magnetic medium can be described by one third-rank nonlinear susceptibility tensor with different components [4], which are either even or odd in magnetization M, describing the crystallographic or the magnetization-induced contribution, respectively:

$$P_i(2\omega, \pm M) = (\chi^+_{ijk}(\pm M) \pm \chi^-_{ijk}(\pm M))E_j(\omega)E_k(\omega). \qquad (1)$$

The odd components $\chi_{ijk}^-$ change sign upon magnetization reversal and therefore give rise to the magnetic asymmetry in the MSHG response.

The measured intensity in a fixed experimental geometry with opposite magnetization direction can be written as a sum of effective tensor components:

$$I^\pm(2\omega) \propto \left|\chi_{eff}^+(2\omega) \pm \chi_{eff}^-(2\omega)\right|^2 \quad (2)$$

where $\chi_{eff}^+$ and $\chi_{eff}^-$ are linear combinations of the even and odd tensor elements and Fresnel factors $\alpha_{ijk}$:

$$\chi_{eff} = \sum_{i,j,k} \alpha_{ijk} \chi_{ijk} \quad (3)$$

The magnetic asymmetry can then be defined as

$$A = \frac{I^+ - I^-}{I^+ + I^-} \quad (4)$$

Because the asymmetry $A$ is normalized to the total SHG intensity, it does not depend on the intensity of the fundamental light. Thus, a large asymmetry A can be expected since the odd and even components may be in the same order of magnitude.

**Table I.** The nonzero elements of the MSHG susceptibility tensor for Fe/AlGaAs (001) structure in the longitudinal configuration ($\vec{M} \parallel \hat{x}$). The surface is in the $\hat{x}$-$\hat{y}$ plane. $\chi^+$ and $\chi^-$ denote the even and odd elements with magnetization. The even elements from interface and bulk are indicated by subscripts a and b.

| Sample Orientation | $\chi^+$ | $\chi^-$ | Input Polarization | Output Polarization |
|---|---|---|---|---|
| $M_x \parallel [100]$ |  | yyy | s | s |
|  | $yxz^b = yzx^b$ | yxx, yzz | p | s |
|  | $zyy^a$ |  | s | p |
|  | $xzx^a = xxz^a$ |  | p | p |
|  | $zxx^a$, $zzz^a$ |  |  |  |
| $M_x \parallel [110]$ |  | yyy | s | s |
|  |  | yxx, yzz | p | s |
|  | $zyy^{a,b}$ |  | s | p |
|  | $xzx^{a,b} = xxz^{a,b}$ |  | p | p |
|  | $zxx^{a,b}$, $zzz^a$ |  |  |  |

Although MSHG is intrinsically sensitive to the interface magnetization in the centrosymmetric layer (Fe: bcc), a bulk response may be generated in the noncentrosymmetric material (AlGaAs: zinc blende) which significantly reduces the nonlinear magneto-optical effect. However, this bulk derived signal can be avoided by judicious selection of input- and MSHG-light polarization combination. To quantitatively describe this, we have analyzed the tensor components $\chi_{ijk}^+$ and $\chi_{ijk}^-$ of Fe/AlGaAs (001) heterostructure, summarized in Table I.

For both principle crystallographic directions, <100> and <110>, the odd magnetic components $\chi_{ijk}^-$ contribute only to the s-polarized MSHG signal. In particular, for s-input polarization and s-polarized MSHG signal, a large bulk response from the GaAs (001) substrate can be avoided since it contains only a magnetization-induced response. This polarization combination yields the following azimuthal dependence of the SHG intensity:

$$I(2\omega,\phi,\pm M) = \left| \pm A^{s,s} \pm B^{s,s} \cos 4\phi \right|. \tag{5}$$

Here, the coefficients $A^{s,s}$ and $B^{s,s}$ are combinations of magnetization-induced nonlinear susceptibility tensor elements and Fresnel factors. These terms change sign upon magnetization reversal. However, this pattern does not yield any effect of magnetization reversal for the MSHG intensity measurement. In addition, the signal in ideal $S_{in}S_{out}$ combination is normally small as shown in SHG measurements from centrosymmetric structures [6], which may lead to very low signal to noise ratio in the M-H loops. Therefore, a mixture of small p-polarized bulk SHG and MSHG signal is required to obtain a good magnetic contrast in such measurements. This can be achieved simply by rotating the analyzer by a few degrees.

Experiments

MSHG experiments are performed with a Ti:sapphire amplifier system (Spectra Physics) generating 150-femtosecond pulses with 1-mJ energy at 1-KHz repetition rate and 800-nm wavelength. The attenuated laser beam (15 mW) is focused to a ~500-μm diameter spot on the sample at an angle of incidence of 45° (Fig. 1). The SHG signal is generated in the direction of the reflected laser beam, and is detected with a high signal-to-noise ratio using a photomultiplier tube (PMT) and a chopper in combination with a lock-in amplifier. The SHG light is effectively filtered using the combination of a prism and blue filters. For measurements of the azimuthal dependence of the MSHG signal, the sample is mounted on a computer-controlled rotation stage between the poles of an electromagnet, with the magnetic field applied in the plane of the Fe film.

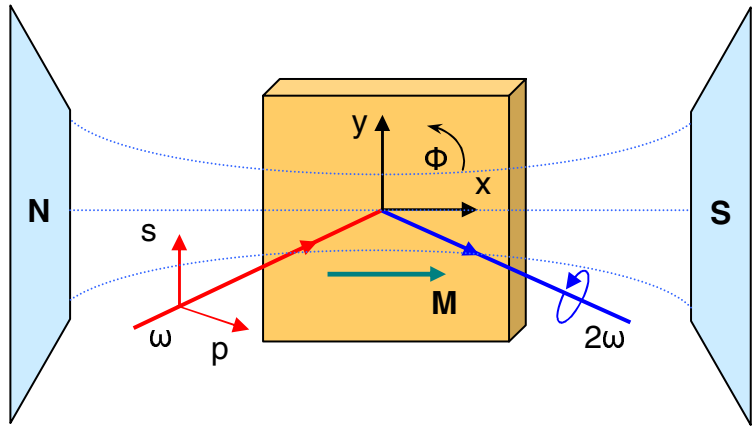

**Figure 1**. Schematic experimental geometry: the sample rotates azimuthally and the magnetic field is applied in the longitudinal magneto-optical configuration. The notations p and s denote the polarizations parallel and perpendicular to the plane of incidence, respectively.

Results and Discussions

Figure 2 shows the rotational MSHG intensity from 10-nm Fe film on AlGaAs (001). In this measurement, the applied field is set at ±600 Oe so that the magnetization is aligned along the field direction. The incident light is s-polarized and the SHG signal passes through the analyzer set at an angle $\alpha = -6°$ with respect to s-polarized direction.

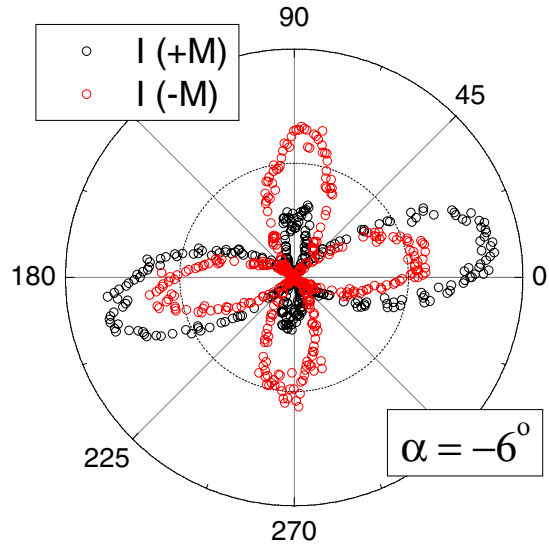

**Figure 2**. Azimuthal dependence of longitudinal nonlinear magneto-optical effect in a 10-nm Fe film on AlGaAs (001). The incident light is s-polarized, and the MSHG signal passes through the analyzer set at an angle $\alpha = -6°$ with respect to s-polarized direction. At $\phi = 0$, M is parallel to the [1-10] axis.

The two-fold symmetry revealed in the rotational MSHG intensity originates from the p-polarized signal generated in the bulk AlGaAs. The isotropic component from the interface also contributes to the p-polarized signal, leading to the one-fold symmetry. However, the pronounced angular shifts of the four loops cannot be described by the even and odd susceptibility components given in Eq. (5). Therefore, one has to take into account additional anisotropic contributions to the second-order nonlinear response. In particular, a vicinal surface and/or a quadrupole-allowed response can lead to such a rotational anisotropy. A detailed discussion of these crystallographic contributions is beyond the scope of this paper.

Despite several non-magnetic contributions to the SHG signal, a large magnetic asymmetry can be clearly seen along all of the sample directions. According to Eq. (4), we plot in Fig. 3 the asymmetry as a function of the analyzer angle α with magnetization along [1-10] direction. We can rewrite the magnetic asymmetry as [7]:

$$A(\alpha) = 2\Phi_k ctg\alpha [1+(\Phi_K ctg\alpha)^2]^{-1} \cos\varphi \quad (6)$$

with the nonlinear complex Kerr angle $\Phi_K e^{i\varphi} = e^{i\varphi} E_s(2\omega)/E_p(2\omega)$.

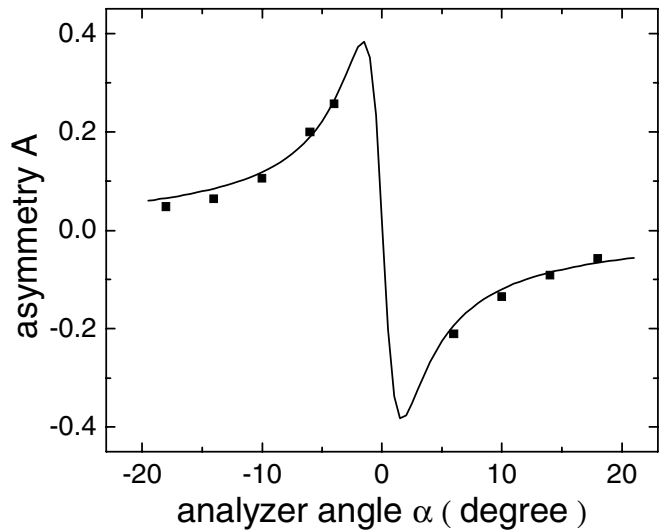

**Figure 3**. Asymmetry A as a function of the analyzer angle α for magnetization along the [1-10] direction in Fe/AlGaAs (001) and s-polarized incident light. Solid line is the best fit according to Eq. (6).

We note in Fig. 3 that the magnetic asymmetry is significantly reduced when *p*-polarized light generated in bulk AlGaAs is mixed-in at large α (>20°). However, the SHG intensity is very small at α<1°, leading to low signal-to-noise ratio and large uncertainty of the asymmetry measurement. Furthermore, the MSHG signal from a 20-nm Fe film is within our experimental error. In this film, the interface contribution can be neglected since the film thickness is larger than the penetration depth of the laser light. This indicates that the oxidized surface of the 10-nm Fe film does not contribute to the MSHG signal. Therefore, in the 10-nm Fe film the large MSHG signal and asymmetry at medium α (4°~20°) must result from the interference between nonlinear optical waves of magnetic and crystallographic origin, generated at the ferromagnetic Fe interface and in bulk AlGaAs, respectively.

We determine the correlated phase φ=66° and the magnitude of the nonlinear Kerr angle $\Phi_K = 1.6°$ from the fit of Eq. (6) to the magnetic asymmetry curve in Fig. 3. In the same geometry, we measured the longitudinal linear Kerr angle of ~0.02°, similar as reported by Doi *et al*. [8]. This corresponds to an enhancement of almost two orders of magnitude for the nonlinear Kerr rotation, thus allowing precise measurements of the interface magnetization reversal process [9].

## Conclusion

This study shows for the first time a large nonlinear magneto-optical Kerr rotation in a non-centrosymmetric ferromagnetic heterostructure, Fe/AlGaAs (001). By judicious selection of the polarization combination for incident laser light and output second-harmonic optical waves, a high magnetic contrast in the MSHG signal can be achieved along all in-plane directions of the anisotropic Fe/AlGaAs (001) structure. Our results demonstrate the high sensitivity of MSHG to interface magnetization in hybrid structures containing non-centrosymmetic materials as well as centrosymmetric multilayers.